\documentclass[journal]{IEEEtran}

 \usepackage{graphicx}
 \usepackage{caption} 
 \usepackage{color}

\usepackage{amsmath}

\usepackage{multirow}

\usepackage{algorithmic}
\usepackage{algorithm}

\begin{document}
\title{Enhancing Blood Glucose Prediction with Meal Absorption and Physical Exercise Information}

\author{Chengyuan Liu, Josep Veh\'{i}, Nick Oliver, Pantelis Georgiou  and Pau Herrero
\thanks{C. Liu, P. Georgiou and P. Herrero are with the Centre for Bio-Inspired Technology, Department
of Electrical and Electronic Engineering, Imperial College London, London SW7 2AZ, UK. E-mail: pherrero@imperial.ac.uk.
 
 J. Veh\'{i} is with Department of Electrical and Electronic Engineering, Universitat de Girona and with CIBERDEM, Girona, 17004, Spain.
 
N. Oliver is with Charing Cross Hospital, Imperial College Healthcare NHS Trust, London W6 8RF, UK.}}

\markboth{}
{Enhancing Blood Glucose Prediction with Meal Absorption and Physical Exercise Information}

\maketitle

\begin{abstract}

Objective: Numerous glucose prediction algorithm have been proposed to empower type 1 diabetes (T1D) management. Most of these algorithms only account for input such as glucose, insulin and carbohydrate, which limits their performance. Here, we present a novel glucose prediction algorithm which, in addition to standard inputs, accounts for meal absorption and physical exercise information to enhance prediction accuracy. 
Methods: a compartmental model of glucose-insulin dynamics combined with a deconvolution technique for state estimation is employed for glucose prediction. In silico data corresponding from the 10 adult subjects of UVa-Padova simulator, and clinical data from 10 adults with T1D were used. Finally, a comparison against a validated glucose prediction algorithm based on a latent variable with exogenous input (LVX) model is provided. 
Results: For a prediction horizon of 60 minutes, accounting for meal absorption and physical exercise improved glucose forecasting accuracy. In particular, root mean square error (mg/dL) went from $26.68\pm 3.58$ to $23.89\pm 3.32$, $p<0.001$ (in silico data); and from $37.02\pm 5.14$ to $35.96\pm 4.65$, $p<0.001$ (clinical data - only meal information). Such improvement in accuracy was translated into significant improvements on hypoglycaemia and hyperglycaemia prediction. Finally, the performance of the proposed algorithm is statistically superior to that of the LVX algorithm ($26.68\pm 3.58$ vs. $32.80\pm4.58$, $p<0.001$ (in silico data);  $37.02\pm 5.14$ vs. $49.17\pm 13.80$ $p<0.01$ (clinical data).  Conclusion: Taking into account meal absorption and physical exercise information improves glucose prediction accuracy.
 
\end{abstract}



\section{Introduction}

\IEEEPARstart{T}{ype} 1 diabetes mellitus (T1DM) is an autoimmune condition characterized by elevated blood glucose levels due to the lack of endogenous insulin production \cite{daneman2006type}. People with T1DM require exogenous insulin delivery to regulate glucose levels. Current therapies for T1DM management require measuring capillary glucose levels several times per day and the administration of insulin by means of multiple daily injections (MDI) or continuous subcutaneous insulin infusion (CSII) with pumps. More recently, the appearance of subcutaneous continuous glucose monitoring (CGM) allows access to virtually continuous glucose concentrations measurements (e.g. every 5 minutes), glucose rate-of-change, and allows their retrospective analysis. In addition, real-time devices include alerts and alarms for concentrations outside of specified ranges and or rapid changes in glucose. Clinical data suggest that CGM can improve overall glucose control, as measured by glycated haemoglobin \cite{lind2017continuous}, and can reduce the burden of extreme glucose values (hypo- and hyperglycaemia) \cite{polonsky2017impact}. In addition, CGM technology has opened the door to new technologies for managing glucose levels such as sensor-augmented insulin pumps with low-glucose insulin suspension \cite{buckingham2010prevention} and the artificial pancreas \cite{trevitt2016artificial}. One important feature of CGM-based technologies is the ability to forecast glucose concentrations in order to avoid undesired events, such as hypoglycaemia and hyperglycaemia, by enabling pre-emptive action (e.g. insulin dose to address hyperglycaemia).

Several glucose forecasting algorithms have been proposed by different authors, with a comprehensive and extensive review being recently published, which provides a taxonomy of the different types of existing algorithms  \cite{oviedo2016review}. In addition, commercial  applications of such technology already exists in the form of sensor-augmented insulin pumps (e.g. Medtronic MiniMed $640$G with Smart Guard) that has been proven to reduce nocturnal hypoglycaemia using predictive glucose alerts and a predictive low-glucose insulin suspension system \cite{buckingham2010prevention}. 
 
Some glucose prediction algorithms use continuous glucose monitoring (CGM) data as the unique source of information to forecast glucose levels while others use additional exogenous inputs such as meal intake and insulin injections, which are know to influence blood glucose levels \cite{oviedo2016review}. Taking such information into account has been proven to improve forecasting accuracy \cite{zhao2012predicting}. Furthermore, additional information such as meal absorption and physical exercise information can potentially further improve accuracy \cite{ferrara2006effects, shah2017effect}.

In this work, we introduce a novel model-based glucose prediction algorithm which uses deconvolution of the CGM signal to estimate some model states in order to improve prediction accuracy. In addition to using CGM data, insulin boluses and carbohydrate intake information, information about meal absorption and physical exercise is taken into account to further enhance prediction accuracy.
For comparison purposes, the latent variable with exogenous input (LVX) algorithm proposed by Zhao et al. has been selected as reference in glucose forecasting since it has showed superiority when compared against existing techniques in the literature and its source code is publicly available \cite{zhao2012predicting}. The proposed algorithm is evaluated for different prediction horizons ranging from 5 to 120 minutes, with special focus on the $60$-minute horizon \cite{oviedo2016review}. Finally, predictive hypoglycaemia and hyperglycaemia prediction capabilities of the tested algorithms are evaluated.

For testing purposes, the UVa-Padova type 1 diabetes simulator (T1DMS) \cite{kovatchev2009silico} was extended in order to include a physical exercise model, a richer meal-model library and intra-day variability. In addition, a two-week clinical dataset from a cohort of 10 adult subjects with T1DM was employed. Finally, the performance of the tested algorithms was evaluated by means of root mean square error (RMSE) and Clarke Error Grid Analysis (EGA).
 
\section{Methods}

The proposed glucose prediction algorithm is based on composite minimal model of glucose-insulin regulation in type 1 diabetes \cite{herrero2012robust} that uses deconvolution of the continuous glucose monitoring (CGM) signal to estimate some of the model states. In particular, the states of the gastrointestinal model are estimated using the technique proposed by Herrero et al. \cite{herrero2012simple}, which has been proved to be a simple but effective way to estimate the glucose rate of appearance from mixed meals. Finally, meal information (i.e., carbohydrate amount and absorption type), insulin boluses and physical exercise are considered as exogenous inputs. Note that compared to the model used in the T1DM simulator, the employed composite minimal model is relatively simple and easy to identify, while providing sufficient complexity to model glucose-insulin dynamics. The effectiveness of such composite model was evaluated by Gillis et al. for glucose prediction using a Kalman filter technique \cite{gillis2007glucose} and by Herrero and associates for detecting faults in insulin pump therapy \cite{herrero2012robust}.

\subsection{Composite minimal model}

The employed composite model of glucose regulation in type 1 diabetes is composed of the minimal model of glucose disappearance proposed by Bergman and colleagues \cite{bergman1979quantitative}, and the insulin and carbohydrate absorption models proposed by Hovorka et al.\cite{hovorka2004nonlinear}. 

\subsubsection{Minimal model of glucose disappearance}
The minimal model of glucose disappearance \cite{bergman1979quantitative} is described by the equations
\begin{align}
\label{1_Gp}
 \dot{G}(t)&=  -(S_G + X(t))G(t) + S_GG_b + \frac{R_a(t)}{VW},\\
\label{2_X}
 \dot{X}(t)&=  -p_2X(t) + p_2S_II(t),
\end{align}
where $G(t)$ is the glucose concentration, $X(t)$ is the insulin action, $R_a$ is the glucose rate of appearance from ingested meals, $I(t)$ is the plasma insulin concentration, $S_G$ is  the fractional glucose effectiveness, $S_I$ is the insulin sensitivity, $p_2$ is the insulin action rate, $G_b$ is the basal glucose, $V$ is the distribution volume, and $W$ is the subject's body weight.

\subsubsection{Insulin absorption model}
The plasma insulin concentration is estimated by means of the subcutaneous insulin absorption model proposed by Hovorka et al. \cite{hovorka2004nonlinear}, which is described by the following equations.
\begin{align}
\dot{S}_1(t)&= u_1(t) - \frac{S_1(t)}{t_{maxI}},\\
\dot{S}_2(t)&=  \frac{S_1(t) - S_2(t)}{t_{maxI}},\\
\dot{I}(t)&=-k_eI(t) + \frac{S_2(t)}{V_it_{maxI}},
\end{align}
where $S_1(t)$ and $S_2(t)$ are the subcutaneous short-acting insulin compartments, $I(t)$ denotes the plasma insulin concentration, the input $u_1(t)$ represents the subcutaneous insulin infusion, $t_{maxI}$ is the time to maximum insulin absorption, $V_i$ is the distribution volume of insulin and $k_e$  is the decay rate.

\subsubsection{Glucose absorption model}
The glucose rate of appearance ($R_a$) is calculated according to the gastrointestinal absorption model by Hovorka et al. \cite{hovorka2004nonlinear}, which is represented by the equations
\begin{align}
\label{6_Ra1}
\dot{R}_{a1}(t)&=-\frac{R_{a1}(t)-A_gu_2(t)}{t_{maxG}},\\
\label{7_Ra}
\dot{R}_a(t)&=-\frac{R_a(t)-R_{a1}(t)}{t_{maxG}},
\end{align}
where
$R_{a1}(t)$ denotes the glucose appearance in the first compartment, $R_a(t)$ represents the rate of glucose appearance, the model input $u_2(t)$ denotes the carbohydrate intake amount, $t_{maxG}$ is the time to maximum glucose rate of appearance and $A_g$ is the carbohydrate bioavailability .

\subsubsection{Physical exercise}
Schiavon and coauthors have showed that physical exercise produces significant changes on insulin sensitivity \cite{schiavon2013postprandial, schiavon2013silico}. Since the effect of physical exercise on glucose uptake and insulin sensitivity is not explicitly modelled within the employed minimal model, its effect is taken into account by modifying the parameter $S_I$, which models the ratio between endogenous glucose production and glucose uptake, during the duration of the exercise. In particular, insulin sensitivity was modified as follows
\begin{align}
\label{SI}
S_I:=\left\{\begin{array}{lr} k_{ex} S_I^o,& \text{during exercise}\\[2mm]
S_I^o,& \text{during resting}\end{array}\right.
\end{align}
where $S_I^o$ is the insulin sensitivity in absence of exercise and $k_{ex}$ is a constant that represents the effect of physical exercise on insulin sensitivity. Note that the employed model of physical exercise only accounts for the short-term effect of anaerobic on glucose levels (i.e. glucose update) and does not account for the long-term effect on insulin sensitivity.
In this work, a $30$-minute exercise at $50\%$ VO2max (see Section \ref{sect:insilico}) has been considered, and $k_{ex}$ was empirically fixed to 3.


\subsubsection{Meal absorption}
\label{sect:absorption}

Meal composition has a profound effect on blood glucose levels \cite{shah2017effect}. Therefore, taking this information into account can potentially enhance glucose forecasting performance. To account for this information in a practical way from the user's perspective, meals were classified as fast, medium and slow absorption. In particular, fast-absorption meals were considered have more than $60\%$ of the area under the curve (AUC) of the rate of glucose appearance ($R_a$) profile appeared within the first two hours since the meal ingestion; a slow-absorption meals to have less than $80\%$ of AUC of $R_a$ profile appeared within four hours, and medium-absorption meal otherwise.
To take meal absorption information into account within the employed glucose absorption model, the time-to-maximum absorption rate $t_{maxG}$ was modified as follows 
\begin{align}
\label{tmaxG}
t_{maxG}:=\left\{\begin{array}{lr} t_{maxG}^o-t_{l},& \text{fast absorption}\\[2mm]
t_{maxG}^o,& \text{medium absorption}\\[2mm]
t_{maxG}^o+t_{d},& \text{slow absorption}\end{array}\right.
\end{align}
where $t_{maxG}^o$ is the default time-to-maximum absorption rate (i.e. medium absorption) for a given subject, $t_l$ and $t_d$ represent the time shift on the time-to-maximum absorption rate due to different meal absorption rates. In particular, $t_l$ and $t_d$ were empirically fixed to $20$ minutes.

\begin{figure}[!t]
     \centering
             \includegraphics[width=3.4in]{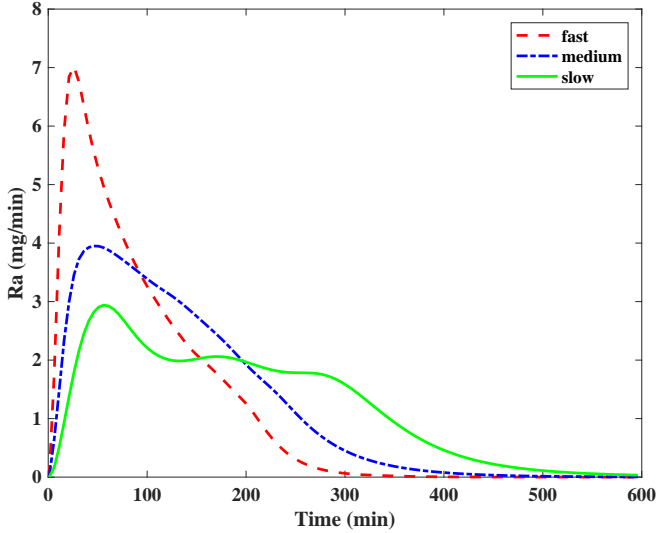}
             \caption{$R_a$ profiles corresponding to the fast, slow and medium meals of the employed UVa-Padova simulator for a 60 grams intake of carbohydrates.}
             \label{fig_Ra}
\end{figure}

Fig. \ref{fig_Ra} shows the average $R_a$ profiles corresponding to the fast, slow and medium meals of the employed UVa-Padova simulator for a 60 grams intake of carbohydrates.

\subsection{Glucose prediction algorithm} \label{PA}

The proposed glucose prediction algorithm uses a discretised version of the presented composite model (Equations \ref{1_Gp}-\ref{7_Ra}). For this purpose, a forward Euler's Configuration with 1-minute step size is used to simulate the model. 

Let $$x(k) = f(x(k-1),p,u(k-1))$$ be the system equations representing a discretised version of the described composite model, where $k$ denotes the sampling instant. Let
$x=\begin{bmatrix} G \!& X \!& S_1 \!& S_2 \!& I \!& R_{a1} \!& R_a \end{bmatrix}$ represent the model states; $p=\begin{bmatrix} k_e \!&\! t_{maxI} \!&\! V_i \!&\!  A_g \!&\!  t_{maxG} \!&\!  S_G \!&\!  p_2 \!&\! W \!&\! V \!&\!  S_I \end{bmatrix}$ represent the model parameters; and $u\!=\! \begin{bmatrix} CHO \!&\! I_B \!&\! E_X \!&\! M_A \end{bmatrix}$ represent the model inputs, where $CHO$ denotes the amount of ingested carbohydrates, $I_B$ denotes the insulin boluses (units), $E_X$ denotes an exercise flag (i.e. true or false) and $M_A$ denotes the meal absorption (i.e. slow, medium, fast).

To improve the forecasting capability of the proposed algorithm, the model states of the gastrointestinal sub-model ($R_a$ and $R_{a1}$) are estimated in real-time (e.g. every 5 minutes)  by doing a deconvolution of the continuous glucose monitoring signal using the technique proposed by Herrero et al. \cite{herrero2012simple}. Such models estates were selected for being highly dependent on meal composition. 
The glucose rate of appearance ($R_a$) in the second compartment is estimated as
\begin{align}
\hat{R}_a(k)=\left[\dot{G}(k)+(S_G+X(k))\bar{G}(k)-S_GG_b\right]VW,
\end{align}
where $\dot{G}$ is the derivative of the glucose measurements calculated as the slope of the linear regression of three consecutive glucose values, $\bar{G}$ is the sensor measurement and $X$ is the insulin action (Equation \ref{2_X}). In order to reduce the influence of the measurement disturbance, the derivative is bounded by $|\dot{G}| \leq 1$ mg/dL per min. To further reduce the effect of sensor noise on the $\hat{R}_{a}$ estimation, a moving average filter is applied, 
\begin{align}
\hat{R}_{a}(k):=\frac{\sum_{i=k-n}^{i=k} R_a(i)+\hat{R}_{a}(i)}{(n+1)},
\end{align}
where $n$ is the length of the moving window ($n=3$).

The glucose appearance in the first compartment is then estimated as 
\begin{align}
\hat{R}_{a1}(k)=\dot{R}_{a}(k) t_{maxG}+\hat{R}_{a}(k),
\end{align}
where $\dot{R}_{a}$ is the derivative of $\hat{R}_{a}$. 

Then, the states $R_a$ and $R_{a1}$ are calculated as a weighted average between the simulated values with the model $f$ and the estimated ones by the deconvolution technique as follows
\begin{align}
R_a(k)&:=Q\hat{R}_{a}(k)+(1-Q)R_a(k),\\[2mm]
R_{a1}(k)&:=Q\hat{R}_{a1}(k)+(1-Q)R_{a1}(k).
\end{align}
where $Q \in [0,1]$ is a tuning parameters that allows putting more weight on the model estimation or on the deconvolution technique. Note that parameter $Q$ allows to decide if more trust is put on the model estimation ($R_a$) or on the estimation using deconvolution ($\hat{R}_{a}$).

Similarly, the plasma glucose state is updated as
\begin{align}
G(k):=Q\bar{G}(k)+(1-Q)G(k).
\end{align}

Finally, the discrete model $f$ is evaluated over the predefined prediction horizon (PH) to obtain the desired forecasted glucose. 

In this works, prediction horizons ranging from 5 to 120 minutes are evaluated (Figure \ref{fig_rmse}). However, special emphasis is put on the  60-minute horizon since it is one of the most employed ones in the literature \cite{oviedo2016review} and is the horizon currently used by the predictive low-glucose insulin suspend (SmatGuard) implemented in the Medtronic MiniMed 640G sensor-augmented insulin pump (Medtronic, Northridge, CA, US).  

\subsection{Model parameter identification} \label{PI}

The proposed glucose prediction algorithm is individualised by identifying some of the model parameters using retrospective data.
Since identification of all model parameters is not possible due to identifiability problems, some of the parameters, which are know to have less inter-subject variability, were fixed to mean populations values (i.e. $S_G$, $V$, $V_i$, $k_e$, $p_2$, $A_g$) \cite{dalla2005insulin}, while others were set by using \textit{a priori} known information from the subjects, such as body weight ($W$) and basal glucose ($G_b$). Finally, parameters $S_I$,  $t_{maxI}$ and $t_{maxG}$ were identified by minimising the mean absolute relative difference (MARD) between the predicted glucose ($G_f$) and the corresponding glucose measurements. Matlab \textit{fmincon} constrained optimisation routine was employed for this purpose. Constraints for the identified parameters were $S_I \in [0.001, 0.005]$ $min^{-1} ~per~ \mu U/ml$, $t_{maxI} \in [50,140]$ $min$ and $t_{maxG} \in [50,140]$ $min$. Table \ref{table_Pa} shows the employed values for the model parameters indicating which ones are \textit{a priori} known and which ones are identified.

\begin{table*}[!t]
\caption{Values of the parameters used in the forecasting algorithm. $*$ indicates parameters that are identified and $**$ parameters that are known from \textit{a priori} information from the subjects. The rest of the parameters are fixed to mean population values obtained from the scientific literature \cite{herrero2012simple,hovorka2004nonlinear}.}
\label{table_Pa}
\centering
\begin{tabular}{|c|c|c|c|c|c|c|c|c|c|c|c|c|c|}
\hline
Parameter & $S_G$ & $S_I$  & $G_b$ & $V$ & $V_i$ & $W$ &  $t_{maxI}$ & $t_{maxG}$ &  $k_e$ & $p_2$ & $A_g$ & $Q$ & $PH$  \\
\hline
Value & $0.02$  & $*$ & $**$ & $0.9$ &  $1.2$ &  $**$ & $*$ & $*$  &  $1.5$ & $0.02$ & $0.85$   & $0.7$ & $30$\\
\hline
Units &$min^{-1}$ &$min^{-1} ~per~ \mu U/ml$ & $mg/dl$ &$dl/kg^2$ & $ml/kg$& $kg$ &  $min$ & $min$ & $min^{-1}$ &$min^{-1}$ & -- & -- & min\\
\hline 
\end{tabular}
\end{table*}

\subsection{\textit{In silico} testing}
\label{sect:insilico}

The latest version of the UVa-Padova T1DM simulator (v3.2) \cite{kovatchev2009silico} was used to evaluate the proposed glucose forecasting algorithm. The 10 available adult subjects were used for this purpose. The open-loop insulin therapy provided by the simulator was employed to generate the datasets. A one-week scenario with a daily pattern of carbohydrate dose intake of $7am$ ($70g$), $13pm$ ($100g$) and $7pm$ ($80g$) ($\pm 20min$) was chosen. The selected CGM and insulin pump models to perform the simulations were the Dexcom G4 and Deltec Cozmo.

Intra-day variability was emulated by modifying some of the parameters of the model described in \cite{dalla2007meal}. In particular, meal variability was emulated by introducing meal-size variability ($CV=10\%$), meal-time variability ($STD=20$) and uncertainty in the carbohydrate estimation (uniform distribution between $-30\%$ and $+20\%$) \cite{BRAZEAU201319}. Meal absorption rate ($k_{abs}$) and carbohydrate bioavailability ($f$) were considered to vary by $\pm 30\%$ and $\pm10\%$ respectively. To account for variability in meal composition, the $33$ available meals in the simulator were considered. Note that each cohort had $11$ different meals (i.e. $10$ individuals plus an average individual). In addition, $16$ mixed meals obtained from clinical data extracted from scientific publications were included. A mixed-meal model library was obtained using the technique for estimating the rate of glucose appearance proposed by Herrero et al. in \cite{herrero2012simple,herrero2013application}. Details about the meal library are provided in Appendix A. By using the absorption classification criteria introduced in Section \ref{sect:absorption}, of the $49$ considered meal, $31$ were classified as fast absorption, $15$ as medium absorption and $3$ as slow absorption.
Intra-subject variability in insulin absorption model parameter ($k_d$, $k_{a1}$, $k_{a2}$) was assumed $\pm30\%$ \cite{haidar2013pharmacokinetics,herrero2017enhancing}. Finally, physical exercise was introduced as described in \cite{schiavon2013silico}. In particular, a $30$-minute exercise $CV=10\%$ at $50\%$ VO2max was considered at $3pm$ ($\pm 20min$).

In order to test the benefit of accounting for meal and exercise information in the glucose predictions, four configurations of the proposed algorithm were considered. These are:
\begin{itemize}
\item $Configuration~1$: exercise and meal type information (i.e. slow, medium, fast) are not taken into account. 
\item $Configuration~2$: only exercise information is taken into account. 
\item $Configuration~3$: only meal type information is taken into account. 
\item $Configuration~4$: both meal type and exercise information are taken into account. 
\end{itemize}

The latent variable model with exogenous input (LVX) algorithm proposed by Zhao et al. \cite{zhao2012predicting} was chosen to compare its performance against the proposed technique.

Finally, in order to train both the proposed algorithm and the LVX algorithm, a one-week training dataset, different from the testing scenario, was employed.

\subsection{Clinical data testing}
\label{sect:clinical}

Although significant intra-day variability was considered in the selected \textit{in silico} scenario, it still cannot be compared to a real-life scenario. In order to test the proposed algorithm with real clinical data, a one-week clinical dataset from the 10 adult subjects with T1DM undergoing a clinical trial evaluating the benefits of an advanced insulin bolus calculator was employed \cite{Reddy2016Clinical}. Since no reliable information about physical exercise and meal composition was available for the clinical dataset, the proposed algorithm was evaluated making the assumption that breakfast is fast absorption and lunch and dinner are medium absorption (i.e. Configuration 3). Not that not having such information in a reliable way might limit the benefits of the proposed algorithm. Finally, the algorithm without considering information about meal absorption and exercise (i.e. $Configuration~1$) and the LVX algorithm were also evaluated and compared. Two different one-week datasets were employed for training and testing purposes.


\subsection{Evaluation metrics}

In order to evaluate the forecasting accuracy of the algorithms, the root mean square (RMSE) and the percentage of values in A-region of the Error Grid Analysis (EGA) were used. RMSE is calculated as
$$RMSE = \sqrt{\frac{\sum_{k=1}^{N}(\hat{G}-\bar{G})^2}{N}},$$
where $\hat{G}$ is the forecasted value, $\bar{G}$ the glucose measurement, and $N$ is the total number of glucose measurements. 
EGA express the clinical significance of the error between the forecasted glucose value and the actual measurement. In particular, the A-region of EGA represent the percentage of the forecasted glucose values that deviate from the actual measurements within the range of $\pm20\%$, or when both the forecasted and the actual measurements indicate hypoglycaemia (i.e. $|\hat{G}-\bar{G}|\leq 20\%\bar{G} $ or $\hat{G}\leq 70mg/dL~with~ \bar{G}\leq 70mg/dL$).
Although other metrics exist to evaluate the clinical significance of the committed error, such as continuous glucose error grid analysis, the EGA is the most widely used one \cite{wentholt2006critical}. 

In addition, the efficiency of predicting hypo- and hyperglycaemia prediction are evaluated by the sensitivity (SEN), specificity (SPC), $F_1$ score, and the Matthews correlation coefficient (MCC). The sensitivity measures the percentage of correct predictions of hypoglycaemia (or hyperglycaemia) events and the specificity measures the percentage of correct prediction within the target range (e.g. $70mg/dL<\bar{G}\leq 180mg/dL$) with the formula
$$SEN = \frac{TP}{TP+FN},~SPC = \frac{TN}{TN+FP},$$
where $TP$ denotes the number of true positives (i.e. correct prediction of hypo-/hyperglycaemia), $FN$ denotes the number of false negatives (i.e. missed prediction of hypo-/hyperglycaemia), $TN$ denotes the number of true negatives (i.e. correct prediction of glucose within target range), and $FP$ denotes the number of false positives (i.e. false prediction of hypo-/hyperglycaemia). 
Finally, two metrics to evaluate the quality of the binary classifications were included: the $F_1$-score and the Mathew's correlation coefficient (MCC). $F_1$ score is calculated as
$$F_1 = \frac{2TP}{2TP+FP+FN},$$
and MCC as
$$MCC\! = \!\frac{TP\cdot TN\!-\!FP\cdot FN}{\sqrt{(TP\!+\!FP)(TP\!+\!FN)(TN\!+\!FP)(TN\!+\!FN)}}. $$
 
\section{Results}

\subsection{\textit{In silico} results}

The distribution of the identified model parameters for the employed \textit{in silico} cohort is $S_I=0.00275\pm0.0014$ $min^{-1} ~per~ \mu U/ml$,  $t_{maxI}=114.6\pm21.6$ $min$ and $t_{maxG}=68.9\pm6.8$ $min$. 

For the 10 virtual adult subjects, Table \ref{table_Rmse} shows: the prediction accuracy expressed as $RMSE$ and the percentage of pairs (i.e predicted vs. measurement) in region A  of the EGA for different prediction horizons corresponding to the four configurations and LVX algorithm.
Table \ref{table_Accuracy} shows the sensitivity, specificity, $F_1$ score and the MCC of hypoglycaemia and hyperglycaemia prediction with a prediction horizon of 60 minutes, corresponding to the four evaluated configurations and the LVX model-based algorithm . Such results are expressed as mean and standard deviation ($Mean\pm STD$) and statistical significance with respect to the row below is indicated with $^*$ for $p<0.001$, $^+$ for $p<0.01$, and $^T$ for $p<0.05$.  

\begin{table*}[!ht]
\caption{RMSE ($mg/dL$) and A-region of the EGA ($\%$) expressed in $Mean \pm STD$ for the four considered configurations and the LVX algorithm corresponding to different prediction horizon ($PH$ in $minutes$) and evaluated on the 10-adult virtual population. The statistical significance with respect to the row below is indicated with $^*$ for $p<0.001$, $^+$ for $p<0.01$ and $^T$ for $p<0.05$}.
\label{table_Rmse}
\centering
\begin{tabular}{|c|c|c|c|c|c|c|c|c|}
\hline
\multirow{2}{*}{Config-}  & \multicolumn{2}{|c|}{$PH = 30$}  & \multicolumn{2}{|c|}{$PH = 60$} &\multicolumn{2}{|c|}{$PH = 90$}  &\multicolumn{2}{|c|}{$PH = 120$}\\
\cline{2-9}
\multirow{2}{*}{uration}   & RMSE  & A-region  & RMSE   & A-region  & RMSE   & A-region   & RMSE   & A-region  \\
                       & ($mg/dL$) & ($\%$) & ($mg/dL$) & ($\%$)  & ($mg/dL$) & ($\%$)  & ($mg/dL$) & ($\%$) \\
\hline
$4$  & $13.04\!\pm \!1.64^*$ & $93.43\!\pm \!1.72^*$ & $23.89\!\pm \!3.32^*$ & $77.86\!\pm \!5.36^*$ & $33.11\!\pm \!4.50^+$ & $66.14\!\pm \!5.75^+$ & $39.14\!\pm \!5.60^*$ & $58.47\!\pm \!5.73^+$\\
\hline
$3$  & $13.79\!\pm \!1.75^*$  & $92.24\!\pm \!1.84$ & $24.84\!\pm \!3.34^*$ & $77.14\!\pm \!5.18^*$ & $33.95\!\pm \!4.54^*$ & $65.74\!\pm \!5.79^*$ & $39.81\!\pm \!5.64^*$ & $58.14\!\pm \!5.66^*$\\
\hline
$2$ &  $13.69\!\pm \!1.66^T$ & $92.63\!\pm \!1.69^*$ & $25.75\!\pm \!3.54^+$  & $74.84\!\pm \!5.66^*$ & $35.54\!\pm \!4.75^*$ & $61.82\!\pm \!6.44^T$ & $42.21\!\pm \!5.82^*$ & $53.58\!\pm \!5.98$\\
\hline
$1$  &  $14.40\!\pm \!1.77^*$ & $91.44\!\pm \!1.85$ & $26.68\!\pm \!3.58^*$  & $74.32\!\pm \!5.52^+$ & $36.38\!\pm \!4.80^*$ & $61.56\!\pm \!6.42^+$ & $42.88\!\pm \!5.86^*$ & $53.39\!\pm \!5.82^*$\\
\hline
$LVX$ & $16.24\!\pm \!2.12$ & $91.37\!\pm \!1.75$ & $32.80\!\pm \!4.58$  & $68.85\!\pm \!6.74$ & $59.09\!\pm \!17.39$ & $40.83\!\pm \!17.28$ & $87.44\!\pm \!27.57$ & $26.09\!\pm \!16.44$\\
\hline
\end{tabular}
\end{table*}

\begin{table*}[!ht]
\caption{Hypoglycaemia and hyperglycaemia prediction results for the 10 virtual adult subjects considering a 60 minutes prediction horizon. Results are expressed as $Mean\pm STD$ and statistical significance with respect to the row below is indicated with $^*$ for $p<0.001$, $^+$ for $p<0.01$ and $^T$ for $p<0.05$.}
\label{table_Accuracy}
\centering
\begin{tabular}{|c|c|c|c|c|c|c|c|c|}
\hline
Config-  &  Hypo SEN & Hypo SPC& Hypo F1 & Hypo MCC &  Hyper SEN & Hyper SPC &  Hyper $F_1$ & Hyper MCC \\
	uration	                   & ($\%$) 		       & ($\%$) & ($\%$) 		       & ($\%$)  & ($\%$)                 & ($\%$) 		       & ($\%$) & ($\%$)\\
\hline
$4$   & $93.30\!\pm\! 1.88^+$ & $97.65\!\pm\! 0.91^+$& $87.20\!\pm\! 4.01^+$ & $85.91\!\pm\! 3.84^+$& $65.88\!\pm\! 9.07^T$ & $96.18\!\pm\! 2.88$& $60.42\!\pm\! 11.99^+$ & $57.75\!\pm\! 9.54^+$\\
\hline
$3$ &  $92.54\!\pm \!1.95^T$  & $97.58\!\pm \!0.87^+$& $86.43\!\pm\! 4.54$ & $85.07\!\pm\! 4.31$& $64.37\!\pm\! 9.90$ & $96.08\!\pm\! 2.92^T$& $59.26\!\pm\! 12.99^+$ & $56.46\!\pm\! 10.69^+$\\
\hline
$2$  & $92.10\!\pm\!2.26^+$ & $97.69\!\pm\!0.96^T$& $86.79\!\pm\! 4.03^+$ & $85.40\!\pm\! 3.91^+$& $61.83\!\pm\! 10.29^T$ & $95.79\!\pm\! 3.21$& $56.99\!\pm\! 13.17^+$ & $53.87\!\pm\! 10.90^+$\\
\hline
$1$ & $91.61\!\pm \!2.22^*$ & $97.64\!\pm \!0.93^+$ & $86.30\!\pm\! 4.22^*$ & $84.85\!\pm\! 4.08^*$& $60.28\!\pm\! 11.72^*$ & $95.68\!\pm\! 3.31$& $55.87\!\pm\! 13.93^*$ & $52.59\!\pm\! 11.85^*$\\
\hline
$LVX$ & $83.39\!\pm \!5.23$ & $93.18\!\pm \!3.19$& $68.17\!\pm\! 11.99$ & $65.68\!\pm\! 10.75$& $44.68\!\pm\! 12.47$ & $95.46\!\pm\! 1.18$& $42.18\!\pm\! 17.09$ & $38.22\!\pm\! 14.82$\\
\hline
\end{tabular}
\end{table*}

\begin{figure}[!t]
     \centering
             \includegraphics[width=3.3in]{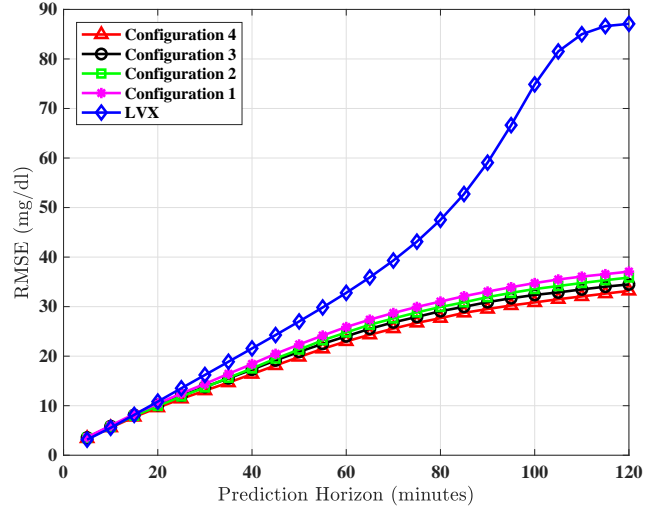}
             \caption{Mean $RMSE$, in $mg/dL$, for the four configurations and the LVX algorithm against to evaluated prediction horizons corresponding to the 10 virtual adults.}
             \label{fig_rmse}
\end{figure}

\begin{figure}[!t]
     \centering
             \includegraphics[width=3.4in]{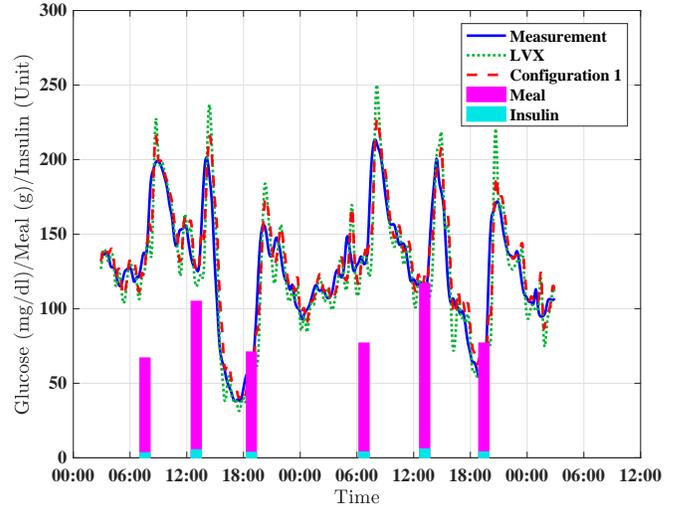}
             \caption{Two-day period close-up of the prediction results for subject adult $1$. The simulated continuous glucose measurements are showed in solid blue line, the prediction results of the LVX method are showed in dotted green line, and the prediction results of the $Configuration~1$ are showed in dashed red line. Vertical pink bars indicate carbohydrate intakes (grams) and vertical light blue bars indicate insulin boluses (units).}
             \label{fig_a3}
\end{figure}

Fig. \ref{fig_rmse} shows the mean RMSE with regard to the prediction horizons for the four considered configurations and the LVX algorithm.
Fig. \ref{fig_a3} shows a two-day period close-up of the prediction results for subject adult $1$ corresponding to $Configuration~1$ and the LVX method. Note that the LVX method tends to overestimate or underestimate glucose values in the peaks and troughs. 

\subsection{Clinical data results}

The distribution of the identified model parameters for the employed 10-adult cohort are $S_I=0.0011 \pm 0.00015$ $min^{-1} ~per~ \mu U/ml$,  $t_{maxI}=78.36 \pm 16.52$ $min$ and $t_{maxG}= 85.23 \pm 24.86$ $min$.

For the 10 adult subjects, Table \ref{table_Rmse_C} presents the RMSE and the A-region  ($Mean \pm STD$) corresponding to $Configuration ~3$, $Configuration ~1$ and $LVX$ method for the different prediction horizons evaluated on the 10 adult subjects. 
Table \ref{table_CAccuracy} shows the sensitivity, specificity, F1 score and the MCC of hypoglycaemia and hyperglycaemia prediction with a prediction horizon of 60 minutes. These results correspond to $Configuration ~3$, $Configuration ~1$ and the LVX model-based algorithm. Results are expressed as $Mean\pm STD$ and statistical significance with respect to the row below is indicated with $^*$ for $p<0.001$, $^+$ for $p<0.01$, and $^T$ for $p<0.05$. 

\begin{table*}[!ht]
\caption{RMSE ($mg/dL$) and A-region of EGA ($\%$) expressed ($Mean \pm STD$) for $Configuration ~3$, $Configuration ~1$ and the LVX algorithm corresponding to different prediction horizon ($PH$ in $minutes$) and evaluated on the 10 adult subjects. The statistical significance with respect to the row below is indicated with $^*$ for $p<0.001$, $^+$ for $p<0.01$ and $^T$ for $p<0.05$}.
\label{table_Rmse_C}
\centering
\begin{tabular}{|c|c|c|c|c|c|c|c|c|}
\hline
\multirow{2}{*}{Config-}  & \multicolumn{2}{|c|}{$PH = 30$}  & \multicolumn{2}{|c|}{$PH = 60$} &\multicolumn{2}{|c|}{$PH = 90$}  &\multicolumn{2}{|c|}{$PH = 120$}\\
\cline{2-9}
\multirow{2}{*}{uration}   & RMSE  & A-region  & RMSE   & A-region  & RMSE   & A-region   & RMSE   & A-region  \\
                       & ($mg/dL$) & ($\%$) & ($mg/dL$) & ($\%$)  & ($mg/dL$) & ($\%$)  & ($mg/dL$) & ($\%$) \\
\hline
$3$   &  $25.06\!\pm \!5.33$       & $83.05\!\pm \!7.51$   & $35.96\!\pm \!4.65^*$     & $71.57\!\pm \!6.60^*$     & $41.84\!\pm \!3.32^*$     & $66.72\!\pm \!4.71^*$     & $44.73\!\pm \!3.26^*$ & $65.25\!\pm \!4.27^+$\\
\hline
$1$    &  $25.40\!\pm \!5.66^*$ & $82.75\!\pm \!7.87$    & $37.02\!\pm \!5.14^+$  & $70.82\!\pm \!6.82^+$ & $43.25\!\pm \!3.67^*$  & $65.91\!\pm \!4.73^*$ & $46.51\!\pm \!3.70^*$ & $64.45\!\pm \!4.23^*$\\
\hline
$LVX$ & $30.00\!\pm \!9.66$     & $81.13\!\pm \!10.67$ & $49.17\!\pm \!13.80$    & $62.55\!\pm \!12.58$  & $64.74\!\pm \!18.21$   & $50.74\!\pm \!12.71$   & $76.94\!\pm \!21.28$ & $43.02\!\pm \!11.78$\\
\hline
\end{tabular}
\end{table*}

\begin{table*}[!ht]
\caption{Hypoglycaemia and hyperglycaemia prediction results for the 10 adult subjects considering a 60 minutes prediction horizon. Results are expressed as mean and standard deviation ($Mean\pm STD$) and statistical significance with respect to the row below is indicated with $^*$ for $p<0.001$ and $^T$ for $p<0.05$.}
\label{table_CAccuracy}
\centering
\begin{tabular}{|c|c|c|c|c|c|c|c|c|}
\hline
Config-  &  Hypo SEN & Hypo SPC& Hypo F1 & Hypo MCC &  Hyper SEN & Hyper SPC &  Hyper $F_1$ & Hyper MCC \\
	uration	                   & ($\%$) 		       & ($\%$) & ($\%$) 		       & ($\%$)  & ($\%$)                 & ($\%$) 		       & ($\%$) & ($\%$)\\
\hline
$3$   & $67.24\!\pm\! 12.56^+$ & $95.42\!\pm\! 3.78$& $49.99\!\pm\! 12.59^+$ & $48.85\!\pm\! 12.86^+$& $86.66\!\pm\! 7.21$ & $88.21\!\pm\! 3.26^T$& $84.36\!\pm\! 9.64$ & $73.92\!\pm\! 8.97^T$\\
\hline
$1$ & $66.26\!\pm \!11.96^+$ & $95.26\!\pm \!3.81^+$ & $48.71\!\pm\! 12.18^*$ & $47.57\!\pm\! 12.32^*$& $86.67\!\pm\! 7.31^*$ & $88.00\!\pm\! 3.23$& $84.28\!\pm\! 9.65^*$ & $73.73\!\pm\! 8.98^*$\\
\hline
$LVX$ & $60.54\!\pm \!14.40$ & $93.79\!\pm \!5.16$& $40.52\!\pm\! 13.90$ & $39.30\!\pm\! 13.70$& $78.17\!\pm\! 12.06$ & $87.26\!\pm\! 3.76$& $79.33\!\pm\! 11.56$ & $65.64\!\pm\! 10.43$\\
\hline
\end{tabular}
\end{table*}

\begin{figure}[!t]
     \centering
             \includegraphics[width=3.3in]{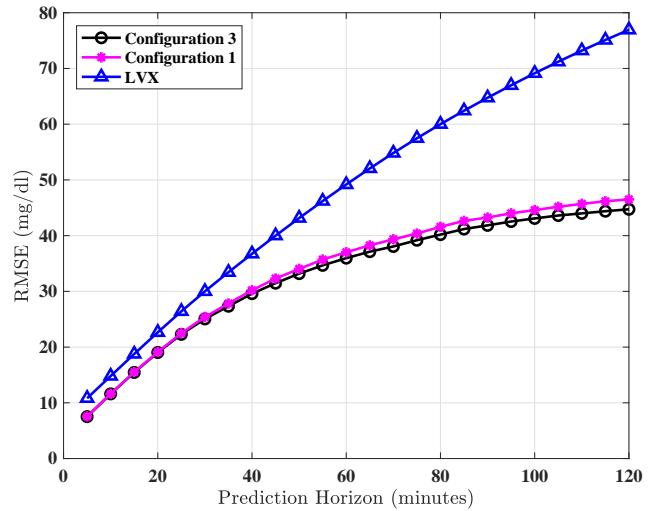}
             \caption{Mean $RMSE$, in $mg/dL$, corresponding to $Configuration ~3$, $Configuration ~1$ and $LVX$ algorithm for different prediction horizons and evaluated on 10 adults subjects.}
             \label{fig_rmse_C}
\end{figure}

\begin{figure}[!t]
     \centering
             \includegraphics[width=3.4in]{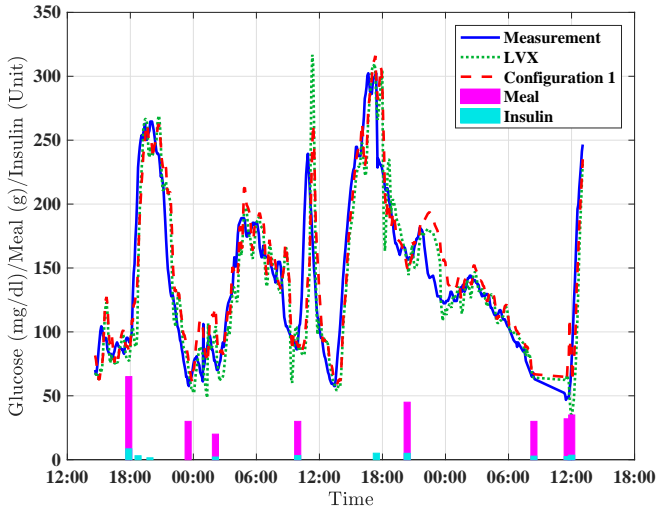}
             \caption{two-day period close-up of the prediction results for $Configuration~1$ and $LVX$ method. The simulated continuous glucose measurements are showed in solid blue line, the prediction results of the LVX method are showed in dotted green line, and the prediction results of the $Configuration~1$ are showed in dashed red line. Vertical pink bars indicate carbohydrate intakes (grams) and vertical light blue bars indicate insulin boluses (units).}
             \label{fig_c1}
\end{figure}

Fig. \ref{fig_rmse_C} shows the mean RMSE for different prediction horizons corresponding to $Configuration ~3$, $Configuration ~1$ and $LVX$ algorithm.
Fig. \ref{fig_c1} shows a two-day period close-up of the prediction results for $Configuration~1$ and $LVX$ algorithm corresponding to a selected subject .

\section{Discussion}

The obtained results show that accounting for information about physical exercise can significantly improve the accuracy of a glucose forecasting algorithm, having the information about meal composition has a bigger impact on the results. However, the major improvement is achieved when both sources of information are taken into account (e.g. RMSE (mg/dL) from  $26.68\pm 3.58$ to $23.89 \pm 3.32$ (in silico data)). 
The improvement in accuracy has a significant impact on hypoglycaemia and hyperglycaemia prediction, having a major impact on the latter one ($F_1$: from $55.87 \pm  13.93\%$ to $60.42\pm 11.99\%$ and MCC: from $52.59 \pm  11.85\%$ to $57.75 \pm  9.54\%$).

Note that accounting for meal absorption requires an additional input by the user. Hence, in addition to standard training carbohydrate counting, people with T1D should also receive training to learn how to classify between slow, medium and faster absorption meals. Also note that physical activity information can be easily gathered using off-the-shelf activity monitors (e.g. Fitbit), but could also be manually entered.

The presented results show that the proposed algorithm outperforms the LVX model-based algorithm proposed by  Zhao et al. \cite{zhao2012predicting} ($26.68\pm 3.58$ vs. $32.80\pm 4.58$). These results are consistent in the both evaluated \textit{in silico} and clinical scenarios. Note that the difference in performance between the two algorithms is much more accentuated for longer prediction horizons. Finally, it is important to remark the significantly smaller standard deviation of the proposed algorithm performs when compared to the LVX algorithm, which seems to indicate that our approach generalises better.

\section{Conclusion and Future Work}

Accounting for information about meal absorption (slow, medium, fast) and physical exercise (duration and intensity) improves the performance of a glucose forecasting algorithm, having the information about meal composition has a bigger impact on the results.
When compared against an existing glucose forecasting algorithm ($LVX$ model), the proposed algorithm, based on a compartmental model of glucose-insulin dynamics combined with a deconvolution technique for state estimation, provides superior performance in terms of prediction accuracy and hypoglycaemia and hyperglycaemia prediction. 
Although the obtained results are conclusive, using longer datasets for training the models could lead to improved results.

The presented algorithm is currently being clinically evaluated as part of the safety system of a mobile-based decision support system for type 1 diabetes management within the framework of the European project PEPPER (Patient Empowerment through Predictive PERsonalised decision support) \cite{PEPPER}.  
Current work to further improve the accuracy of the proposed algorithm include accounting for insulin sensitivity circadian variations and a physical exercise model to account for the long term effect of exercise on insulin sensitivity.


\appendices
\section{Mixed-meal model library}
\label{Appendix_B}

To build the employed mixed-meal model library within the UVa-Padova T1DM simulator \cite{kovatchev2009silico}, the scientific literature was reviewed for clinical trials studying the effect of meal composition on non-diabetic subjects, which included mean population plasma glucose and plasma insulin concentration data, meal composition, and body weight. In addition, the duration of the trial needs to be long enough to allow glucose and insulin levels at the end of the trial to return to basal conditions and the sampling rate needs to be high enough to capture glucose and insulin dynamics. Data from 16 mixed meals fulfilling the above criteria were found in scientific publications for healthy subjects. Table \ref{table_Mm} shows the information for each of the selected mixed meals, average weight of the studied subjects and the corresponding bibliographic reference. 

To estimate the rate of glucose appearance ($R_a$) corresponding to the chosen meals, a simple technique for estimating $R_a$ proposed and validated by Herrero and colleagues was employed \cite{herrero2012simple}. The employed technique, which is based on the glucose-insulin minimal model, only requires the identification of the insulin sensitivity from the minimal model, since it is based on the hypothesis that the rest of the model parameters can be considered to vary in relatively small ranges. This hypothesis originates from the experimental evidence that inter-subject variability of these parameters is not very large \cite{DallaManE954}.
\begin{table*}[!t]
\caption{Mixed meals information and bibliographic references.}
\label{table_Mm}
\centering
\begin{tabular}{|c|p{7.8cm}|c|c|c|c|}
\hline
Meal \# &Ingredients &Weight (Kg) &CHO (g) &CHO; Prot.; Fat (\% energy) &Reference\\
\hline
$1$ & Scrambled eggs, Canadian bacon, Gelatin (Jell-O) & $77$ & $77$ & $45;~15;~40$ & \cite{dalla2005insulin} \\
\hline
$2$ & White bread, low-fat cheese, sucrose, oil, butter & $82.3$ & $111$ & $55;~15;~30$ & \cite{galgani2006acute}\\
\hline
$3$ & Fat Milk, white rice, low-fat cheese, fructose, pear, bran-cookies, oil &$82.3$ & $112.3$ & $55;~15;~30$ & \cite{galgani2006acute}\\
\hline
$4$ & Pasta, oil (low fat) & $57$ & $75$ & $80;~15.4;~4.6$ & \cite{normand2001influence}\\
\hline
$5$ & Pasta, oil (medium fat) & $57$ & $75$ & $56;~10.8;~33.2$ & \cite{normand2001influence}\\
\hline
$6$ & Pasta, oil (high fat) & $57$ & $75$ & $37.4;~7.2;~55.4$ & \cite{normand2001influence}\\
\hline
$7$ & Rice, pudding, sugar and cinnamon & $65^{\star}$ & $50.5$ & $74.6;~14.2;~11.2$ & \cite{freckmann2007continuous}\\
\hline
$8$ & Toast, honey, ham, curd cheese, orange juice & $65^{\star}$ & $50.2$ & $26.2;~16.5;~56.7$ & \cite{freckmann2007continuous}\\
\hline
$9$ & Pear barley & $59.8$ & $50$ & $79;~15;~5$ & \cite{brand2005glycemic}\\
\hline
$10$ & Instant mashed potato & $59.8$ & $50$ & $78;~5.5;~4.5$ & \cite{brand2005glycemic}\\
\hline
$11$ & $2$ slices of bread, $1$ and $\frac{1}{2}$ eggs, $1$ tea spoon of margarine and orange juice & $65$ & $50$ & $49;~22.3;~28.7$ & \cite{edes1998glycemic}\\
\hline
$12$ & Cereal, coconut, chocolate, fruit and whipping cream & $76^{\star}$ & $93$ & $18;~16;~66$ & \cite{koutsari2000postprandial}\\
\hline
$13$ & Oats, coconut, almonds, raisins, honey, sunflower oil, banana, double cream and milk & $61.9$ & $121.1$ & $48.6;~6.9;~48$ & \cite{whitley1997metabolic}\\
\hline
$14$ & Same as meal $13$ & $61.9$ & $70.3$ & $28.2; ~6.6; ~65.2$ & \cite{whitley1997metabolic}\\
\hline
$15$ & Same as meal $13$ & $61.9$ & $50$ & $20;~6.1; ~73.9$ & \cite{whitley1997metabolic}\\
\hline
$16$ & Oat loop cereal, milk, white bread, margarine, strawberry jam, orange juice & $67^{\star}$ & $68.8$ & $57;~19;~24$ & \cite{wolever2003long}\\
\hline
\multicolumn{6}{p{8cm}}{$\star$ Estimated from BMI.}
\end{tabular}
\end{table*}

Then, the estimated $R_a$ profiles were fitted to the the gastrointestinal model of the UVa-Padova T1DM simulator \cite{dalla2006system}, which equations are described below.
\begin{align}
\dot{q}_{sto1}(t)&=-k_{21}q_{sto1}(t)+D\delta (t), \\
\dot{q}_{sto2}(t)&=-k_{empt}(t)q_{sto2}(t)+k_{abs}q_{sto1}(t), \\
\dot{q}_{gut}(t)&=-k_{abs}(t)q_{gut}(t)+k_{empt}(t)q_{sto2}(t), \\
\dot{R}_{a}(t)&=-fk_{abs}(t)q_{gut}(t),
\end{align}
where, $q_{sto1}$ and $q_{sto2}$ are the amounts of glucose in the stomach (solid and liquid phase, respectively), $k_{21}$ is the rate of grinding in the stomach, $\delta$ is the impulse function, $D$ is the amount of ingested glucose, $q_{gut}$ is the glucose mass in the intestine, $k_{abs}$ is the rate constant of intestinal absorption, $R_a$  is the glucose rate of appearance in plasma, $f$ is the fraction of the intestinal absorption which actually appears in plasma and $k_{empt}$ is the rate of gastric emptying, which is represented by a nonlinear function describing a slow down of glucose emptying rate and later recovery, based on available physiological knowledge and which depends on the total amount of glucose in the stomach as follows:
\begin{align}
\nonumber
\dot{k}_{empt}(t) &= k_{min}+\frac{k_{max}-k_{min}}{2} \left\{ \tanh  [ \alpha (q_{sto}(t)-bD) ]  \right. \\
&\quad - \left.\tanh [ \beta (q_{sto}(t)-cD)] +2 \right\},\\
\dot{q}_{sto}(t)& = q_{sto1}(t)+q_{sto2}(t), \\
\alpha & =  \frac{5}{2D(1-b)},\\ 
\beta & =  \frac{5}{2Dc}, 
\end{align}
where $k_{min}$ and $k_{max}$ are the minimal and maximal absorption rates respectively, $b$ is the percentage of the dose $q_{sto}$ for which $k_{empt}$  decreases at $(k_{max} - k_{min})/2$ and $c$ is the percentage of the dose $q_{sto}$ for which $k_{empt}$ is back to $(k_{max} - k_{min})/2$. 

Table \ref{table_Mparame} show the identified gastrointestinal model parameters corresponding to the $16$ mixed meals presented in Table\ref{table_Mm}. Finally, Figure \label{fig:Ra} show the curve fitting of 16 estimated $R_a$ profiles to the selected gastrointestinal model.

\begin{table*}[!t]
\caption{Gastrointestinal model parameters corresponding to the 16 selected mixed meals. Coefficient of variation (\%) provided by the Matlab \textbf{lsqnonlin} optimization routine is reported in brackets.}
\label{table_Mparame}
\centering
\begin{tabular}{|c|c|c|c|c|c|}
\hline
Meal \# & $k_{min}$ & $k_{max}$ & $k_{abs}$  & $b$ & $d$\\
\hline
$1$ & $0.0123~(1.0)$ & $0.0575~(2.7)$ & $0.0388~(4.5)$ & $0.6947~(1.0)$ & $0.0145~(5.3)$ \\
\hline
$2$ & $0.0128~(1.0)$ & $0.0176~(1.1)$ & $1.4807~(4.8)$ & $0.9905~(0.2)$ & $0.4388~(3.2)$\\
\hline
$3$ & $0.0110~(1.5)$ &$0.0207~(4.8)$ & $0.0779~(13.2)$ & $0.9593~(0.7)$ & $0.1633~(4.4)$\\
\hline
$4$ & $0.0096~(1.3)$ & $0.0155~(1.5)$ & $0.0719~(6.0)$ & $0.7834~(0.9)$ & $0.3303~(3.2)$\\
\hline
$5$ & $0.0025~(44.5)$ & $0.0128~(0.9)$ & $0.1276~(8.7)$ & $0.7729~(0.5)$ & $0.6081~(4.7)$\\
\hline
$6$ & $0.0675~(14.3)$ & $0.0256~(13.5)$ & $0.0332~(25.4)$ & $0.3917~(4.6)$ & $0.7238~(3.5)$\\
\hline
$7$ & $0.2000~(28.5)$ & $0.0396~(2.2)$ & $0.0104~(1.8)$ & $0.4048~(7.1)$ & $0.4955~(2.4)$\\
\hline
$8$ & $0.0129~(3.9)$& $0.0224~(22.2)$ & $0.0273~(45.7)$ & $0.7375~(8.5)$ & $0.1964~(27.8)$\\
\hline
$9$ & $0.0098~(1.7)$ & $0.0335~(5.9)$ & $0.0509~(11.8)$ & $0.8031~(0.9)$ & $0.1979~(3.2)$\\
\hline
$10$ & $0.0326~(2.8)$ & $0.1422~(6.5)$ & $0.0355~(3.8)$ & $0.9003~(1.1)$ & $0.0547~(7.5)$\\
\hline
$11$ & $0.0199~(1.8)$ & $0.0896~(7.6)$ & $0.0374~(5.5)$ & $0.9224~(0.6)$ & $0.0939~(5.7)$\\
\hline
$12$ & $0.0115~(0.2)$ & $0.0200~(0.6)$ & $1.6872~(1.5)$ & $0.8578~(0.4)$ & $0.3289~(1.5)$\\
\hline
$13$ & $0.0138~(0.7)$ & $0.0201~(0.9)$ & $0.0946~(3.6)$ & $0.8443~(0.5)$ & $0.3668~(2.3)$\\
\hline
$14$ &  $0.0102~(0.4)$ & $0.0211~(0.6)$ & $1.6260~(1.6)$ & $0.9939~(0.04)$ & $0.3283~(0.6)$\\
\hline
$15$ &  $0.0110~(0.8)$ & $0.0196~(0.5)$ & $0.1563~(4.8)$ & $0.9982~(0.02)$ & $0.4613~(1.0)$\\
\hline
$16$ & $0.0104~(2.3)$ & $0.0215~(1.0)$ & $2.9750~(3.9)$ & $0.9028~(0.4)$ & $0.5045~(2.3)$\\
\hline
\end{tabular}
\end{table*}

To consider the model parameters identification satisfactory, the following conditions were required to hold, where the operator $\bigtriangleup$ denotes the absolute difference between the reference and predicted Ra profiles for the corresponding metric:
\begin{itemize}
\item	Peak value: $\bigtriangleup Ra_{peak}$ $\leq$ 0.3 $mg \cdot min^{-1} \cdot kg^{-1}$
\item	Peak time: $\bigtriangleup T_{peak}$  $\leq$20 min
\item	Area-under-the-curve: $\bigtriangleup AUC$ $\leq$ 30 \%
\item	Root mean square error (RMSE): RMSE $\leq$ 0.5 $mg \cdot min^{-1} \cdot kg^{-1}$
\end{itemize}
Furthermore, the coefficient of variation (CV) provided by the \textit{lsqnonlin} optimization routine was required to be $CV<50\%$  and the coefficient of determination ($R^2$) to be above $0.8$. Table \ref{table_Metrics} show defined metrics for the evaluated $R_a$ profiles.

\begin{table*}[!t]
\caption{Metrics to evaluate the $R_a$ model fitting to reference $R_a$ profiles.}
\label{table_Metrics}
\centering
\begin{tabular}{|c|c|c|c|c|c|}
\hline
Meal	&	$\bigtriangleup Ra_{peak}$	&	$\bigtriangleup T_{peak}$	&	$\bigtriangleup AUC$	&	RMSE	&	$R^2$ \\

 	&	$mg \cdot min^{-1} \cdot kg^{-1}$	&	$min$	&	$\%$	&	$mg \cdot min^{-1} \cdot kg^{-1}$	&	-  \\
\hline
1	&	0.16	&	1	&	2	&	0.1521	&	0.995 \\
\hline
2	&	0.27	&	7	&	1.6	&	0.31623	&	0.978 \\
\hline
3	&	0.20	&	12	&	1.6	&	0.3458	&	0.952 \\
\hline
4	&	0.14	&	10	&	1.2	&	0.17675	&	0.979 \\
\hline
5	&	0.04	&	12	&	3.4	&	0.2535	&	0.964 \\
\hline
6	&	0.20	&	10	&	12.5	&	0.37702	&	0.818 \\
\hline
7	&	0.22	&	6	&	3.6	&	0.230776	&	0.970 \\
\hline
8	&	0.25	&	16	&	4.3	&	0.17117	&	0.972 \\
\hline
9	&	0.22	&	5	&	5.1	&	0.26335	&	0.918 \\
\hline
10	&	0.17	&	5	&	1.1	&	0.2928	&	0.991 \\
\hline
11	&	0.01	&	3	&	4.0	&	0.25221	&	0.987 \\
\hline
12	&	0.21	&	2	&	2.0	&	0.28166	&	0.984 \\
\hline
13	&	0.21	&	9	&	0.5	&	0.27094	&	0.993 \\
\hline
14	&	0.19	&	8	&	1.3	&	0.11685	&	0.995 \\
\hline
15	&	0.12	&	6	&	2.5	&	0.09367	&	0.995 \\
\hline
16	&	0.07	&	1	&	0.02	&	0.3495	&	0.969 \\
\hline
\end{tabular}
\end{table*}


\section*{Acknowledgment}
This project has received funding from the European Union's Horizon 2020 research and innovation programme under grant agreement 689810, and by the Spanish Ministry of Science and Innovation under Grant DPI2016-78831-C2-2-R.

\bibliographystyle{IEEEtran}
\bibliography{IEEEabrv,IEEEexample}

\end{document}